\documentclass[11pt]{article}
\pdfoutput=1
\usepackage{jheppub}

\makeatletter
\def\@fpheader{\relax}
\makeatother

\usepackage{amstext,amsmath,amssymb,amsfonts,bbm, amsthm}
\usepackage{graphicx}
\usepackage{hyperref}
\usepackage{cases}
\usepackage{tocvsec2}

\settocdepth{subsection}

\def\beq{\begin{equation}}
\def\be{\begin{equation}}
\def\ee{\end{equation}}
\def\bea{\begin{eqnarray}}
\def\eea{\end{eqnarray}}
\def\eeq{\end{equation}}

\def\u{\underline}

\begin{document}

\title{Static pure Lovelock black hole solutions with horizon topology 
${\bf S^{(n)} \times S^{(n)}}$}

\author[1]{Naresh Dadhich\note{nkd@iucaa.ernet.in}}
\author[2]{and Josep M Pons,\note{pons@ecm.ub.edu}}

\affiliation[1]{Centre for Theoretical Physics, Jamia Millia Islamia, New Delhi~110025, India, and 
Inter-University Centre for Astronomy and Astrophysics, Post Bag 4, Pune 411 007, 
India}

\affiliation[2]{Departament d'Estructura i Constituents de la Mat\`eria and 
Institut de Ci\`encies del Cosmos (ICC\-UB), Facultat de Física, Universitat de Barcelona, Diagonal 647, 
E-08028 Barcelona, Catalonia, Spain.}

\vspace{1em}

\date{\small\today}

\abstract{
It is well known that vacuum equation of arbitrary Lovelock order for static spacetime ultimately 
reduces to a single algebraic equation, we show that the same continues to hold true for pure Lovelock 
gravity of arbitrary order $N$ for topology ${\bf S^{(n)} \times S^{(n)}}$. We thus obtain pure 
Lovelock 
static black hole solutions with two sphere topology for any order $N$, and in particular we study 
in full 
detail the third and fourth order Lovelock black holes. It is remarkable that thermodynamical 
stability of black hole discerns between odd and even $N$, and consequently between negative 
and positive $\Lambda$ and it 
favors the former while rejecting the latter.}


\maketitle
\flushbottom

\section{Introduction}
\label{intr}
For higher dimensional considerations of gravity, Lovelock generalization is the most natural and 
appropriate as equation of motion continues to remain second order despite Lagrangian being 
homogeneous polynomial in Riemann. This property singles out Lovelock from all other generalizations 
and it includes Einstein gravity in linear order $N=1$ where $N$ is order of Lovelock 
polynomial action. It is well known that Einstein gravity is kinematic in $d=2N+1=3$ dimension which 
means Ricci 
equal to zero implies Riemann zero. In other words, Riemann is entirely determined in terms of Ricci. 
Could this property be true for all $N$;i.e. does there exist a higher order generalization of Riemann 
such that in all odd 
$d=2N+1$ dimensions it is entirely determined by its Ricci? The answer is yes \cite{Dadhich:2012cv} for pure 
Lovelock gravity for which the action contains only one $N$th order term, besides of course the 
cosmological constant $\Lambda$,  
in the action (there is no sum over $N$). We define \cite{Dadhich:2008df} $N$th order analogue of Riemann 
(trace of its Bianchi derivative yields analogue 
of Einstein tensor and agrees with the one that is obtained by varying $N$th order Lovelock action) 
which is a polynomial in Riemann and its contraction giving corresponding Ricci, and so we have 
Lovelock Riemann and Ricci tensors. Then it turns out that whenever Lovelock Ricci vanishes in odd 
$d=2N+1$ dimensions, so does Lovelock Riemann. That is, pure Lovelock gravity which includes Einstein 
gravity for $N=1$, is kinematic in all odd $d=2N+1$ dimensions. It is Lovelock kinematic because it 
is Lovelock Riemann that vanishes and not the usual Riemann. Kinematicity so defined is thus a 
universal gravitational property in higher dimensions, and if it is to be respected always, for a 
given $N$ there are only two pertinent dimensions $d=2N+1, 2N+2$. For $d<2N+1$, it is vacuous while 
for $d>2N+2$, it is next order $(N+1)$th that should be considered else kinematicity property will be 
violated. We shall therefore adhere to universality of this property and shall 
emphasize the pure Lovelock case for $d=2N+1, 2N+2$.  

Spherically symmetric static black holes in Gauss-Bonnet (GB) gravity have been 
studied \cite{BD, Wheeler:1985qd} and in fact for Lovelock 
gravity in general \cite{Wheeler:1985nh,Whitt:1988ax}. It turns out that ultimately one needs only 
to solve an 
algebraic polynomial of degree $N$.  To manage this polynomial and for other physical considerations, 
it was assumed \cite{Banados:1993ur,Crisostomo:2000bb} that all coupling constants are given in terms 
of the unique ground state, $\Lambda$.
This made the equation degenerate and hence could be trivially solved. On the other hand, if we 
consider 
pure Lovelock equation which has only one $N$th order term with $\Lambda$, then the equation is 
derivative 
degenerate and again could be solved trivially \cite{dpp1}. In all this, black hole horizon was a 
sphere of constant curvature. Some authors have considered more general cases with arbitrary 
couplings for each Lovelock term, see \cite{Camanho:2011rj,Maeda:2011ii}, though an exact solution 
for polynomial equation cannot be obtained beyond $N=4$.

The next generalization came through consideration of horizon topology 
being a product of two spheres. Now horizon space is non-maximally
symmetric Einstein space which has non-zero Weyl curvature. However
both Weyl and Riemann have vanishing covariant derivative. Its first example 
was given by Nariai metric \cite{nariai}. An interesting GB black hole solution with two sphere  
topology was obtained by Dotti and Gleiser \cite{Dotti1} and subsequently its various features like 
uniqueness and stability were studied by several authors \cite{Zegers,Dotti:2008pp, Dotti2, Maeda}. 
It turns out that 
non-zero Weyl curvature makes a non-trivial contribution in solution in terms of a constant which 
is negative for GB case. This implies that $\Lambda$ should always be present and positive and there 
also occurs non central singularity. To avoid that, a range is prescribed for black hole mass in 
terms of $\Lambda$ \cite{Pons:2014oya}.\\     

In this paper our main aim is to extend this framework to pure Lovelock of arbitrary order. 
As a matter of fact, our study goes beyond pure Lovelock case ($d=2N+2$) with horizon topology 
${\bf S^{(N)} \times S^{(N)}}$ to include any $d = 2 d_0 +2,\, d_0\geq N$ with horizon topology
${\bf S^{(d_0)} \times S^{(d_0)}}$. That is, we solve Lovelock $\Lambda$-vacuum equation of arbitrary 
order $N$ for a static black hole with two sphere topology in a spacetime with even dimension 
$d \geq 2 N+2$. It is remarkable that algebraic character of ultimate equation is carried through 
for this enlarged framework as well. The product of two spheres produces solid angle deficits which 
cancel out each-other's contribution in Ricci tensor but not in Riemann. For $N>1$, there is always 
Riemann present in the equation and to cancel out solid angle deficit contribution to it, a constant, 
say $p$, is required which is zero for Einstein $N=1$ case. Further it turns out that 
$\Lambda$ and $p$ always bear opposite sign, the former is negative and positive according to 
$N$ being even and odd, and vice-versa for $p$. Whenever $N$ is even,  $p<0$, $\Lambda$ must be 
present and positive. To avoid non central singularity and for existence of horizon, black hole 
mass has to lie in a range prescribed in terms of $\Lambda$. On the other hand for odd 
$N$ when $p>0$, there is no such constraint. It is remarkable that the similar discerning feature 
also emerges from thermodynamical stability considerations. It favors odd $N$ and $\Lambda<0$ as 
against even $N$ and $\Lambda>0$. It is remarkable that two sphere topology distinguishes between 
odd and even $N$ and consequently negative and positive $\Lambda$ quite clearly. This is a new 
realization that has come up only when we have considered both odd and even $N>1$. 

The paper is organized as follows: In next Section we set up equation of motion for pure Lovelock 
$\Lambda$-vacuum of arbitrary order $N$ for a static metric with two sphere topology, and obtain 
general black hole solution. It is followed by consideration of particular cases for $N=1, 2, 3, 4$. 
In Sec. IV we compare two sphere black holes with the corresponding one sphere ones while Sec. V is 
devoted to thermodynamical considerations. We end up with discussion.

\section{Equation of motion} 
\label{algsol}
Consider static metric configuration with product horizon space, $\bf S^{(d_0)}\times S^{(d_0)}$ 
having symmetry group  $\bf SO(d_0+1)\times SO(d_0+1)$,
\begin{equation}
ds^2 = -A(r) \,dt^2 + \frac{1}{A(r)}\,dr^2 + {r^2}(d S^2_{(d_0)}+ d S^2_{(d_0)})\,,
\label{configA}
\end{equation}
describing an even-dimensional spacetime with dimension $d=2d_0+2$, where $dS^2_{(d_0)}$ is metric of 
$d_0$ dimensional unit sphere. We compute Riemann tensor for the metric and its nonvanishing 
components are as follows:
\bea
D(0,1) &=& R_{0\,1}\! {}^{\!0\,1} = -\frac{1}{2} A''(r)\nonumber\\
D(0,i) &=& R_{0\,i}\! {}^{\!0\,i}= -\frac{1}{2 r}A'(r)\nonumber\\
D(0,i')&=& R_{0\,i'}\! {}^{\!0\,i'}= -\frac{1}{2 r}A'(r)\nonumber\\
D(1,i) &=& R_{1\,i}\! {}^{\!1\,i}= -\frac{1}{2 r}A'(r)\nonumber\\
D(1,i')&=&R_{1\,i'}\! {}^{\!1\,i'}=-\frac{1}{2 r}A'(r)\nonumber\\
D(i,j) &=& R_{i\,j}\! {}^{\!i\,j}=\frac{1}{r^2}(1-A(r))\nonumber\\
D(i',j')&=&R_{i'\,j'}\! {}^{\!i'\,j'}=\frac{1}{r^2}(1-A(r))\nonumber\\
D(i,i') &=& R_{i\,i'}\! {}^{\!i\,i'}=-\frac{1}{r^2}A(r)
\label{theDterms}
\eea
where indices $0, 1$ refer to coordinates $t, r$, and indices $i, i'$ to angle coordinates of two 
spheres. 
Notice that there is a crucial difference, in fact the only one, between the one-sphere case, 
$d S^2_{(d-2)}$, and our present two-spheres case, which is the last entry, $D(i,i')$, 
taking indices from two spheres. All the other components are the same for both configurations.

\vspace{5mm}
For this pure Lovelock Lagrangian of order $N$, we will show that the equation of motion (EOM) for the 
metric (\ref{configA}) is ultimately given by a single algebraic equation. Lovelock Lagrangian of 
arbitrary order $N$ is given by 
\beq
{\cal L}_N=\sqrt{-g}\, 
\delta^{\mu_1\dots \mu_{N}\nu_1\dots \nu_N}_{\rho_1\dots \rho_{N}
\sigma_1\dots \sigma_N} \, 
 R_{\mu_1\nu_1} \!{}^{\!\rho_1\sigma_1}\dots  R_{\mu_N\nu_N}\! {}^{\!\rho_N \sigma_N}- 
 \sqrt{-g} \tilde\Lambda\,,
 \eeq
which leads to EOM 
\beq
[{\cal L}_N]^\mu_\rho=\sqrt{-g}\, 
\delta^{\mu\mu_1\dots \mu_{N}\nu_1\dots \nu_N}_{\rho\rho_1\dots \rho_{N}
\sigma_1\dots \sigma_N} \, 
 R_{\mu_1\nu_1} \!{}^{\!\rho_1\sigma_1}\dots  R_{\mu_N\nu_N}\! {}^{\!\rho_N \sigma_N}- 
 \sqrt{-g}\,\delta^{\mu}_{\rho}\, \tilde\Lambda =0\,,
 \label{lovelockEOM}
\eeq
where $\tilde\Lambda$ is, up to some combinatorial factor, the cosmological constant. 
We keep the factor $\sqrt{-g}$ for reasons of later convenience. \\

Our aim is to implement Eq. (\ref{theDterms}) into EOM (\ref{lovelockEOM}) 
to get a differential equation for $A(r)$. It is obvious that for non-vacuous EOM, $\mu =\rho$, 
and we thus write $[{\cal L}_N]^\mu_\mu =:[{\cal L}_N]_\mu$. From Eq. (\ref{theDterms}), it is clear 
that 
\beq
[{\cal L}_N]_1=[{\cal L}_N]_0
\label{eq0}
\eeq
whereas for angles $\mu=i, i'$  
\beq
[{\cal L}_N]_i=[{\cal L}_N]_{i'} =:[{\cal L}_N]_{{}_{i}}\,, \forall\, i,\,i'\,.
\label{eqi}\eeq
So we are left with two equations, $[{\cal L}_N]_0 =0$ and $[{\cal L}_N]_{{}_{i}} = 0$. With some 
manipulations, which we leave for Appendices A and B, it can be shown that the former is in 
fact a total derivative (Appendix A) while the latter takes the form (Appendix B),
\beq
[{\cal L}_N]_{{}_{i}}=\frac{r}{2 d_0} \partial_r \Big([{\cal L}_N]_0\Big).
\label{app}\eeq
An equation of this kind is indeed expected on the basis of Noether  
identities for gauge theories. Note that here the gauge freedom is general covariance. 
In the case $N=1$ Noether identity Eq.(\ref{app}) is the contracted Bianchi identity for 
Einstein tensor. Thus we need to consider only the equation $[{\cal L}_N]_0=0$\,.

What is truly noteworthy is that $[{\cal L}_N]_0$ is a total derivative. It is well known that this is 
already the case with one sphere topology 
\cite{Wheeler:1985nh,Whitt:1988ax}. It is remarkable that this interesting feature carries forward 
also to the present case of two sphere topology (See Appendix A). Now the $\Lambda$-free term in 
Eq.(\ref{lovelockEOM}) for $[{\cal L}_N]_0$ has the form 
$\frac{\partial}{\partial r}\Big(r^{2 d_0 - 2N + 1} P_N(A(r))\Big)$, where $P_N(A(r))$ 
is a polynomial of degree $N$. Here the $r^{2 d_0}$ factor has its origin in the density $\sqrt{-g}$. 
Then the equation,  $[{\cal L}_N]_0=0$ from Eq. (\ref{lovelockEOM}) becomes 
\beq
\frac{\partial}{\partial r}\Big(r^{2 d_0 - 2N + 1} P_N(A(r))\Big) =  r^{2 d_0}\tilde\Lambda\,,
\label{eom-A}
\eeq 
which trivially integrates (with the appropriate redefinition 
$\displaystyle \Lambda=\frac{\tilde\Lambda}{2 d_0+1}$) to give
\beq
P_N(A(r)) = \Lambda\, r^{2 N} + \frac{M}{r^{2 d_0 -2 N+ 1}}\,,
\label{sol-eom-A}
\eeq
where $M$ is an integration constant. Note that in asymptotic expansion, the potential due to 
$M$ will go as $M/r^{2d_0-1} = M/r^{d-3}$, which is the Schwarzschild potential in $d$ dimension. 
This formally indicates that $M$ is mass of the configuration. This is a general feature of pure 
Lovelock gravity that even though EOM is free of Einstein term yet the potential due to mass approaches asymptotically the Schwarzschild potential 
in the corresponding dimension \cite{dpp2}.

As emphasized earlier, it is very remarkable that the general solution can now be obtained by simply 
solving the above algebraic equation of degree $N$.
 
\vspace{4mm}

The polynomial $P_N(A(r))$ takes the following form, 
\beq
P_N(A) := \sum_{l=0}^N 
\Big(\frac{d_0 !}{(d_0-l)!}\Big)^2\,
C[N-l,d_0-l] \,(-A)^l (1-A)^{N-l}\,.
\label{polin}
\eeq
with the combinatorial object
\beq
C[m,s] = \frac{1}{2^m} \sum_{k=0}^m {m\choose k} \frac{s !}{(s-2 m + 2 k) !} \frac{s !}{(s- 2 k) !} \,.
\label{theC}
\eeq
(An extension of the factorial to negative numbers is understood in this formula (\ref{theC}), 
using the gamma function. In particular notice that $C[N,N]$ vanishes for odd $N$)

This polynomial can also be expressed alternatively as 
\bea
P_N(A) &=&\sum _{l=0}^N \sum _{k=0}^{N-l} \frac{1}{2^{N-l}}\,\frac{N!}{k!\, l!\,(-k-l+N)!}\nonumber \\ 
&&\frac{d_0!}{(d_0-2 k-l)! }\,
\frac{d_0!}{(d_0+2 k-2 (N-l)-l)!}(-A(r))^l(1-A(r))^{N-l}.
\label{polinbis}
\eea   

\vspace{4mm}

Let us write  
\beq
A(r)=\frac{d_0-1}{2 d_0-1}(1-\Psi(r))\,,
\label{subAPsi}
\eeq
and define $Q_N(\Psi)=P_N(A)$.
With this substitution the algebraic equation (\ref{sol-eom-A}) becomes
\beq
Q_N(\Psi):=\Lambda r^{2 N} + \frac{M}{r^{2 d_0 -2 N+ 1}}\,,
\label{polQ}
\eeq 
The substitution (\ref{subAPsi}) has the specific property that the polynomial $Q_N(\Psi)$ is now 
free of $\Psi^{N-1}$ term. This means that sum of its $N$ roots $\Psi_l, l=1,\cdots N$ is zero.  

In terms of $\Psi$, the metric is written as
\begin{equation}
ds^2 = -(1-\Psi(r)) \,dt^2 + \frac{1}{1-\Psi(r)}\,dr^2 
+\frac{d_0-1}{2 d_0-1} {r^2}(d S^2_{(d_0)}+ d S^2_{(d_0)})\,.
\label{configPsi}
\end{equation}
Eq.(\ref{polQ}) is the general algebraic equation for pure Lovelock black hole of arbitrary order 
$N$. Next we consider specific cases for $N = 1, 2, 3, 4$ to show that the known Einstein and 
Gauss-Bonnet black holes \cite{BD, Wheeler:1985qd} are all included in this equation. 
\section{Particular cases}
\label{partcases}
\subsection{Case {\bf $\bf N=1$}: Einstein black hole}
In this case, Eq. (\ref{eom-A}) takes the form 
$$
r^{2 d_0-1}\,\Psi(r)=\Lambda r^{2 d_0 +1} + M\,,
$$ 
giving 
\beq
\Psi(r) =  \Lambda r^2 + \frac{M}{r^{2 d_0-1}}
\eeq
for the metric (\ref{configPsi}). This is the usual Schwarzschild-dS/AdS solution except for constant 
factor $\frac{d_0-1}{2 d_0-1}$ before two spheres metric. Note that this factor causes solid angle 
deficit for each sphere but the two together conspire to cancel out each-other in Ricci giving rise 
to $\Lambda$-vacuum. This feature was already noticed by \cite{Kol:2002xz}. However this cancellation 
cannot carry forward to higher order Lovelock simply because analogue of Ricci or Einstein tensor for 
higher order Lovelock would also involve Riemann. Their contributions to Riemann do not however 
cancel out each-other. For that an additional constant would be required for $N>1$ solutions as as 
we would see in the cases that follow. 

\subsection{Case {\bf $\bf N=2$}: Gauss-Bonnet black hole}

In Gauss-Bonnet case, the substitution (\ref{subAPsi}) turns out to be very efficient, making the 
polynomial (\ref{polin}) proportional to $\displaystyle \Psi^2(r) + \frac{d_0}{(d_0-1)^2 (2 d_0-3)}$. 
Then the solution for Eq. (\ref{eom-A}) is of the form
\beq
\Psi(r) = \pm \sqrt{p + \Lambda r^4 + \frac{M}{r^{2 d_0 -3}}}\,.
\label{n-two-sol}
\eeq
where 
\beq
p=-\frac{d_0}{(d_0-1)^2(2 d_0 -3)}
\eeq
We choose the upper positive sign in solution (\ref{n-two-sol}) for existence of black hole horizon as 
well as for gravity being attractive. Note here the constant $p$  
under the radical is negative and it is the one that compensates the term coming from square of Riemann
in GB Ricci analogue which is quadratic in Ricci and Riemann. As argued earlier for Einstein case that 
solid angle deficits of the two spheres cancel out each other to leave $\Lambda$-vacuum intact while in
the case of GB, there is also a quadratic term in Riemann, it is that which requires this additional 
constant. For pure GB black hole, we have $d=2N+2=6, d_0=2$ and  $p=-2$. This case has been 
comprehensively discussed in \cite{Pons:2014oya}. That is $\Lambda$ and $p$ would always bear 
opposite sign and they are respectively positive and negative for even and odd $N$ and vice-versa. 
We shall verify this feature in the following cases for $N=3, 4$.  

\subsection{Case {\bf $\bf N=3$}: Third order pure Lovelock black hole}
\label{case3}
With the appropriate redefinitions, always preserving the signs of $\Lambda $ and $M$, 
Eq. (\ref{sol-eom-A}) is now, 
\beq
a\, \Psi (r)^3+b\, \Psi (r)-p = \Lambda r^{6} + \frac{M}{r^{2 d_0 -5}}
\label{Q3}
\eeq
with 
\bea
a&=& (d_0-1)^3 (2 d_0-5) (2 d_0-3)>0\,,\nonumber\\
b&=&3 (d_0-1) d_0 (2 d_0-5)>0\,,\nonumber\\ p&=& 2 d_0 (2 d_0+1)>0\,.
\eea
The critical dimension (pure Lovelock) for $N=3$ is $d=2N+2=8,\, d_0=3$, and so we write 
(after redefining $\Lambda $ and $M$ to absorb a factor of $6$)
$$
4\, \Psi (r)^3+3\, \Psi(r)= 7+ \Lambda r^{6} + \frac{M}{r}\,.
$$
For existence of horizon, $\Psi(r_h)=1$, we need $\Lambda r_h^{6} + \frac{M}{r_h}=0$ which gives 
$M=-\Lambda r_h^{7}$.  The equation then becomes 
\beq
4\, \Psi (r)^3+3\, \Psi(r)= 7+ \frac{\Lambda}{r}(r^7 -r_h^7)\,.
\label{eqn4}
\eeq
It is clear that $4\, \Psi^3+3\, \Psi$ is monotonically increasing. To keep $\Psi <1$ (i.e, $A(r)>0$), 
which in Eq.(\ref{eqn4}) means RHS $<7$, we need 
$\displaystyle\frac{\Lambda}{r}(r^7 -r_h^7)<0$. For $r>r_h$, this 
means that $\Lambda$ must be negative $\Lambda<0$ which also ensures $M>0$ in 
$\displaystyle r_h= \Big(\frac{M}{-\Lambda}\Big)^{\frac{1}{7}}$. The solution $\Psi (r)$ takes the form 
$$
\Psi (r)=\frac{\left(\sqrt{S^2+1}+S\right)^{2/3}-1}{2 \left(\sqrt{S^2+1}+S\right)^{1/3}}\,,
$$
with $S= 7- \frac{|\Lambda|}{r}(r^7 -r_h^7)$\,. Note that 
$r>r_h \Rightarrow S< 7 \Rightarrow\Psi (r)<1$. 
This describes the region outside black hole horizon. 
Thus there is no constraint on horizon radius and mass is given by $M=\vert\Lambda\vert r_h^{7}$.

The limit $r_h\to 0$ ($\Rightarrow M\to 0$) is regular for $\Psi(r)$ but then the background 
develops a naked singularity at $r=0$. In fact the Kretschmann scalar, 
$K=R_{\mu\nu\rho\sigma}R^{\mu\nu\rho\sigma}$, behaves for $r\to 0$ as 
\beq
K \sim \frac{24}{r^4}\,.
\label{K}
\eeq
Note also that in the $r\to \infty$ limit, $\Psi (r)\to -\frac{|\Lambda|^{1/3}}{2^{2/3}}r^2$. As 
mentioned earlier, $p=7>0$ and $\Lambda$ is required to be negative. On the other hand for $N=2$ GB 
case, 
it was the reverse, $p=-2<0$ and $\Lambda>0$. That is, $\Lambda$ has always to be non-zero, 
positive for even $N$ and negative for odd $N$ while for $N=1$ it could be both positive and negative. 
Pure Lovelock critical dimension for $N=1$ is $d=4$ for which $d_0=1$ that makes 
$S^{(1)} \times S^{(1)}$
space flat. Therefore dimension has always to be $>4$ and hence Einstein case in this setting is not 
a proper pure Lovelock critical dimension case. We will also see in the next case of $N=4$ that 
$\Lambda>0$ and $p<0$.   

\subsubsection{The noncritical case $d_0>3$}
The $N=3$ case, for $d_0>3$ (non pure Lovelock) allows for a window of positive $\Lambda$. 
Although it is not the case that really interests us (which is $d_0=N=3$), we will discuss it here for 
the shake of completeness.

Let us assume $d_0=4$, which is representative of the general $d_0>3$ case. Equation (\ref{Q3}) is
(an irrelevant positive factor has been eliminated from the definition (\ref{polQ}))
\beq
Q_3(\Psi(r)) =  \Lambda r^{6} + \frac{M}{r^{3}}
\label{polQ1}
\eeq
What happened in the critical, pure Lovelock case was that at the horizon ($\Psi(r_h)=1$) the 
polynomial 
$Q_3$ vanished (the reason is that $P_N(0) =0$ for odd $d_0=N$). Now instead we have $Q_3(1) >0$. 
and the 
polynomial is monotonically increasing, $Q_3'(\Psi) >0$. 

At the horizon we have $Q_3(1) \,(= P_3(0)) = \Lambda r_h^{6} + \frac{M}{r_h^{3}}$. Thus we can get 
$M = P_3(0) r_h^{3} -\Lambda r_h^{6}$ and so we write (\ref{polQ1})) as
\beq
Q_3(\Psi(r)) =  \Lambda ( r^{6} - \frac{r_h^{9}}{r^{3}}) + P_3(0)\frac{r_h^{3}}{r^{3}}\,.
\label{polQ2}
\eeq
Since $\Psi(r_h)=1$, for $r>r_h$ but close enough to the horizon, we need $\Psi(r)<1$, otherwise $A(r)$ 
turns negative. Using that the 
polynomial $Q_3(\Psi)$ is monotonically increasing, this means that we need, at the horizon, 
$\frac{d}{d r} Q_3(\Psi(r))_{|_{r_h}} <0$ which, in terms of the rhs of (\ref{polQ2}), yields the 
condition 
$$\Lambda < \frac{ P_3(0)}{3 r_h^{6}}\,.
$$
Thus there is a window for positive $\Lambda$. Of course any negative value for $\Lambda$ 
is admissible and in such case the analysis continues along the same path as the critical $d_0=N=3$ case. 
The case with positive $\Lambda$ can be parametrized by taking $\Lambda = \frac{ P_3(0)}{(3+a) r_h^{6}}$ 
with $a>0$. We then find that there is a minimum $r_0$ for $\Psi(r)$ ($\Psi'(r_0)=0$) 
at $r_0 = (1+\frac{a}{2})^\frac{1}{9} r_h > r_h$.
Then, for $r>r_0$  $\Psi'(r)$ turns positive and we obtain a second, cosmological deSitter-like horizon at
$r_c$ when $\Psi(r_c)=1$.

\subsection{Case {\bf $\bf N=4$}: Fourth order pure Lovelock black hole}
Again with the appropriate redefinitions of $\Lambda $ and $M$, Eq. (\ref{sol-eom-A}) takes the 
form.
$$
a\, \Psi(r)^4+b\, \Psi(r)^2+c\,\Psi(r)-p = \Lambda r^{8} + \frac{M}{r^{2 d_0 -7}}
$$
with 
\bea
a &=& (d_0-1)^4 (2\, d_0-7) (2\, d_0-5) (2\, d_0-3)>0\,,\nonumber\\
b &=& 6 (d_0-1)^2 d_0 (2\, d_0-7) (2\, d_0-5)>0\,,\nonumber\\
c &=& -8 (d_0-1) d_0 (2\, d_0-7) (2\, d_0+1)<0\,,\nonumber\\
p &=& - 3\, d_0 (d_0 (10\, d_0+3)+2)<0\,.
\eea
All cases for $d_0\geq 4$ are similar and we focus on the pure Lovelock $N=4, d_0=4$ case. Redefining 
again for convenience $\Lambda $ and $M$, the equation for $\Psi(r)$ becomes
\beq
F(\Psi(r))\equiv 45 \Psi(r)^4+24 \Psi(r)^2-32\Psi(r) =-\frac{232}{3} +\frac{\Lambda}{8} r^{8} 
+ \frac{M}{r}\,.
\label{FPsi}
\eeq
\begin{figure}
\includegraphics[width=8cm,angle=0]{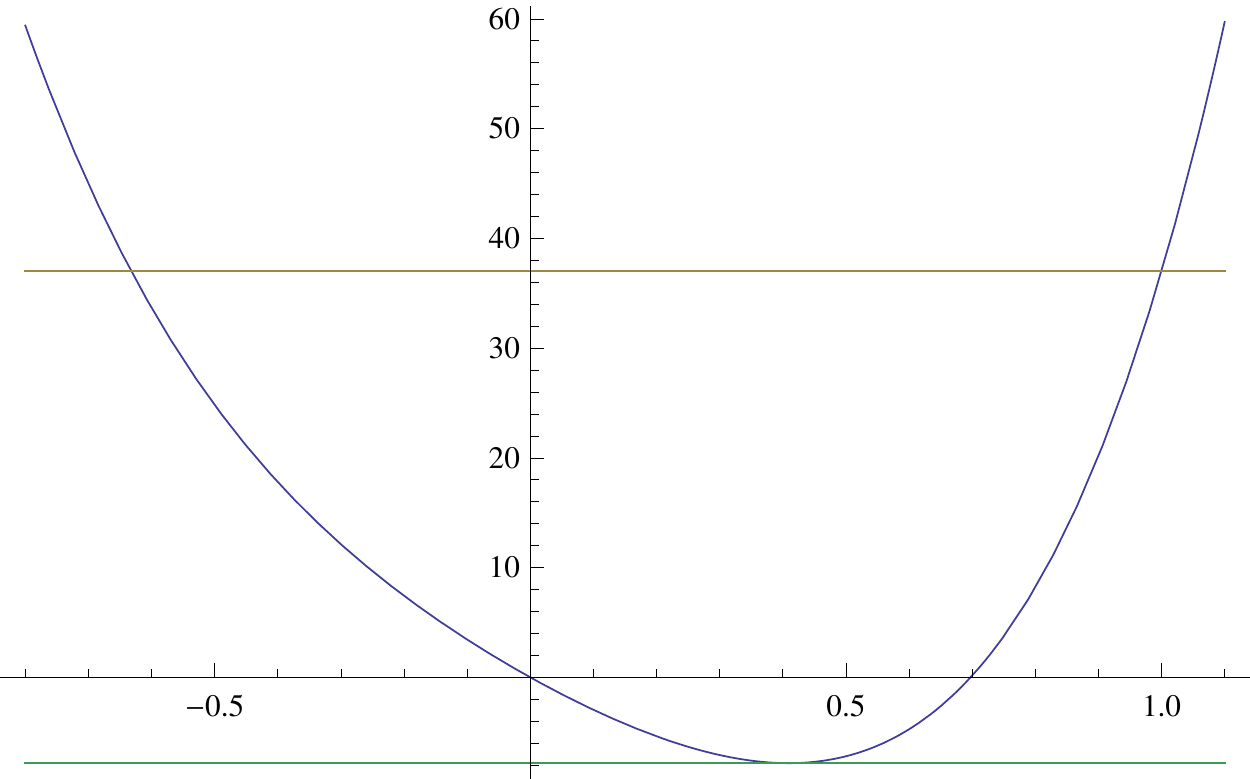}
\caption{Plot of $F(\Psi)$ (Eq. (\ref{FPsi})). \newline
Horizontal lines are\ $y=F(1)=37$ and\ $y=F(\Psi_0) \simeq -7.81$ 
at the minimum $\Psi_0 \simeq 0.41$. \newline 
------------------------------------------------------------------------------------------------} 
\label{L4pol}
\end{figure}
The plot of $F(\Psi)$ against $\Psi$ (see Fig. \ref{L4pol}.) has a parabola like shape with a 
minimum at $\Psi_0 \simeq 0.41$, 
where $F(\Psi_0)$ takes the approximate value $F(\Psi_0) \simeq -7.81$. 
We need to find a region for $r$ for which $\Psi(r)<1$. Note that 
$F(1)=37$. Thus we need values of $r$ such that, approximately
$$
-7.81 <-\frac{232}{3} +\frac{\Lambda}{8} r^{8} + \frac{M}{r}< 37\,,
$$
or
\beq
69.2 <\frac{\Lambda}{8} r^{8} + \frac{M}{r}< 114.3\,.
\label{L4bounds}
\eeq 
Clearly this inequality requires $\Lambda>0$ while $M$ is always to be positive. Otherwise one 
of the bounds on $r$ will be an unwanted naked singularity, $\Psi(r)$ turning complex.  
In fact the minimum for $r$ should respect the range given in (\ref{L4bounds}). 
The minimum is at $r_0=\Big(\frac{M}{\Lambda}\Big)^\frac{1}{9}$, and 
requiring that it satisfies the bounds (\ref{L4bounds}), we end up with following range for 
$\Lambda$ and $M$, 
\beq
61.8 <M^\frac{8}{9}\Lambda^\frac{1}{9} < 101,6\,.
\label{L4boundsML}
\eeq
For $\Lambda$ and $M$ satisfying this bound, we can find a region $r_h<r<r_c$ 
such that $\Psi(r_h)=\Psi(r_c) = 1$ where $r_h$ is black hole and $r_c$ cosmological horizon (similar 
to deSitter), and $r_0$ also satisfies $r_h<r_0<r_c$. Note that here $p=-232/3$ and $\Lambda>0$ 
bearing out the general pattern as envisaged earlier.

\subsection{The general pattern}

As argued in Introduction, the distinguishing property of pure Lovelock gravity is its kinematicity 
in odd $d+2N+1$ dimension. For this property to be universal;i.e. true for all odd $d=2N+1$ 
dimensions, the equation should be valid only in the two critical odd and even dimensions, 
$d=2N+1, 2N+2$. 
We shall therefore stick to the critical, pure Lovelock case, with $d_0=N$. Based on the cases $N=2, 3, 4$ 
and more (we have explored up to $N=12$  but nothing prevents 
from going to any higher $N$), there emerges the following general pattern that $\Lambda$ and $p$ 
(note that up to a positive factor which will be absorbed in $\Lambda$ and $M$, $p$ is defined as 
$p= -Q_N(0)$) always bear opposite sign, the former is respectively positive and negative 
for even and odd values of $N$ while the opposite is true for the latter. 
For $N$ even the shape of $Q_N(\Psi)$ is parabola with a single minimum $\Psi_m$ satisfying
$0<\Psi_m<1$, similar to the case $N=4$ as shown in in  Fig. \ref{L4pol}.
For $N$ odd $Q_N(\Psi)$ is a monotonically increasing function. 

It may be noted that the parameter $p$ arises from countering the deficit angle contribution in the 
Riemann tensor and hence its sign and value is determined by this geometric feature. On the other hand 
for even $N$, $\Lambda>0$ is required for the polynomial (\ref{polQ}) to have a minimum 
while for odd $N$, for some $r$, $Q_N(\Psi)<C[N,N]=0$ in the critical $d_0=N$ dimension, 
and for which $\Lambda$ has to be negative.  

As expected whenever $\Lambda>0$ ($N$ even), there exist both black hole and cosmological 
deSitter-like horizons, and a range of values for black hole mass in terms of $\Lambda$. 
When $\Lambda<0$ ($N$ odd), there is only black hole horizon without any constraint on mass (See Sec. \ref{therm} for non pure Lovelock cases). As a matter of fact, since $ C[N,N]=0$ for $N$ odd, we will have $Q_N(1)=0$ for odd $d_0=N$ because $Q_N(1) = P_N(0)=C[N,d_0]$. Thus it is both BH and cosmological horizons in $d=6, 10, 14, ...$ and only one BH horizon in $d=8, 12, 16, ...$ dimensions. 

\section{Comparison with one sphere case}
\label{one-sph}
It is well known that static spherically symmetric vacuum equation for the general  Lovelock 
Lagrangian ultimately reduces to an algebraic equation \cite{Wheeler:1985nh,Whitt:1988ax}. 
The difference between 
one and two spheres configurations is the absence of the terms with $i'$ indices for the former in 
(\ref{theDterms}). This fact greatly simplifies the analysis and it follows on the same lines as in 
our case. 
The analogue of of the polynomial (\ref{polin}) becomes just a monomial in $(1-A)$, namely
$P_N(A)^{(1)} :=  \frac{(d-2)!}{(d-2 N-2)!} (1-A)^{N}$. The combinatorial factor is customarily 
absorbed in the general Lovelock Lagrangian  
${\cal L}_{\rm Lov.} = \sum_N c_N {\cal L}_N$ with a redefinition of coefficients $c_N$. 
Writing $A(r) = 1-r^2 B(r)$, we have the analogue of EOM (\ref{eom-A}) as 
\beq
\frac{\partial}{\partial r}\Big(r^{2 d_0 + 1} B(r)^N\Big) =  r^{2 d_0} \Lambda\,.
\label{eom-one-sph}
\eeq
If the Lagrangian is a sum of several Lovelock terms $\displaystyle \sum_{N=1}^{N_0} c_N {\cal L}_N$, 
then $B(r)^N$ in 
(\ref{eom-one-sph}) is substituted by $\sum c_N B(r)^N$ and we have, on 
 integrating  (\ref{eom-one-sph}),
\beq \sum_{N=0}^{N_0} c_N B(r)^N =  \frac{M}{r^{2 d_0 + 1}}
\label{sol-one-sph}
\eeq
where the coefficient $c_0$ is proportional to $-\Lambda$ and the integration constant $M$ is 
proportional to  black hole mass. 

This nice simplification in Eq. (\ref{sol-one-sph}), due to the fact that the power of the variable $r$ 
in the bracket in the lhs of (\ref{eom-one-sph})  does not depend on $N$, is no longer 
available in the two-spheres case because of the different structure of the polynomial.
The crucial difference in the two cases is the presence of terms of the type $D(i,i')$ in 
Eq.(\ref{theDterms}) which, 
unlike the terms $D(i,i)$ and $D(i',i')$, are not proportional to $1-A(r)$, but to $A(r)$.

\section{Thermodynamics}
\label{therm}

Even without knowing solution to Eq. (\ref{sol-eom-A}) explicitly, it is possible nevertheless to 
perform thermodynamical analysis for its black hole (BH) solutions 
by just using EOM relying on the fact that these equations are of polynomial 
form. Therefore the method described below can be applied whenever EOM are of this type.
Here we assume that solution $A(r)$ exist with a horizon $A(r_h)=0$, for some range of 
parameters $\Lambda$ and $M$. As a matter of fact we already know that such 
BHs exist for $N=1, 2$ and we have shown in Sec. \ref{partcases} that this is indeed the 
case for arbitrary $N$.

Let us consider Eq. (\ref{sol-eom-A}) for a putative solution $A(r,M)$ (since $\Lambda$ plays no 
special 
role here we do not need to explicitate the dependence of $A$ on $\Lambda$). Thus we have the identity
\beq
P_N(A(r,M)) = \Lambda r^{2 N} + \frac{M}{r^{2 d_0 -2 N+ 1}}\,,
\label{sol-eom-AM}
\eeq
and next we differentiate this identity w.r.t. $M$ and write 
We get
$$
P_N'(0)\, \frac{\partial A}{\partial M}|_{A=0}= \frac{1}{r^{2 d_0 -2 N+ 1}}\,.
$$
We also note that $P_N'(0)<0$ because the definition (\ref{polin}) gives
$$
P_N'(0) = -N \left(d_0^2\, C[N-1,d_0-1] + C[N,d_0]\right)<0\,.
$$
Thus
\beq
\frac{\partial A}{\partial M}|_{A=0} = -\frac{1}{|P_N'(0)|\, r_h^{2 d_0 -2 N+ 1} }\,.
\label{derAM}
\eeq
On the other hand, we can trade $r_h$ and $M$ for each other, which defines the 
function $M(r_h)$ so that $A(r, M(r))=0$ while varying $r$. Thus we can write the identity
$$
A(r,M(r))=0 
$$
as implicitly defining the function $M(r)$. Its $r$-derivative gives the identity 
$$
A'(r,M(r)) + \frac{\partial A}{\partial M}|_{A=0}\,M'(r)=0\,,
$$
where $A'(r,M(r))$ is the derivative with respect to the first argument $r$. In fact the Euclidean method
identifies $A'(r,M(r))$ as the Hawking 
temperature $A'(r,M(r))= 4\pi T(r) $. Thus we have
\beq
\frac{M'(r)}{T(r)} = - 4\pi \frac{1}{\frac{\partial A}{\partial M}|_{A=0}}
= 4\pi |P_N'(0)|\, r^{2 d_0 -2 N+ 1}\,.
\label{MprimT}
\eeq

\vspace{4mm}

\subsection{Entropy}

The entropy is then obtained by integration of the First Law
\begin{equation}
S =\int\frac{d M}{T} = \int_0^{r_h}\frac{M'(r)}{T(r)} dr 
= \frac{4\pi |P_N'(0)|}{2 d_0 -2 N+ 2} r_h^{2 d_0 -2 N+ 2}\,,
\label{entr}
\end{equation}
where we have assumed the vanishing of the entropy when the horizon shrinks to zero. 

Since we only know that $M$ is proportional to the mass of the BH, there is an undetermined numerical 
factor not included in (\ref{entr}). 
The relevant finding nonetheless is that $S$ is proportional to $r_h^{2 d_0 -2 N+ 2}$. Note that in the 
critical Lovelock dimension, $d=2N+2$, the entropy becomes proportional to $r^2$ or, since the area 
${\bf A}$ of the horizon is proportional to $r_h^{2 d_0} = r_h^{2 N}$, it turns out that 
$$
S_{\!{}_{(d_0=N)}}\simeq {\bf A}^{\frac{1}{N}}\,,
$$ 
which confirms that we are in the same universality class as in the spherical pure Lovelock BH, 
as discussed in \cite{dpp1}.

\vspace{4mm}

\subsection{Temperature}

The temperature con be obtained using the identity (\ref{sol-eom-AM}), in which $M$ is a 
fixed parameter. The $r$-derivative of (\ref{sol-eom-AM}), written at the horizon $r_h$ gives
$$
P_N'(0)\, A'(r_h,M) =2 N\, \Lambda r_h^{2 N-1} -(2 d_0 -2 N+ 1) \frac{M}{r_h^{2 d_0 -2 N+ 2}}\,,
$$
which means that, since $P_N'(0)<0$ and $A'(r,M(r))= 4\pi T(r)$,
$$ 
T =\frac{1}{4\pi|P_N'(0)|} \Big((2 d_0 -2 N+ 1) \frac{M}{r_h^{2 d_0 -2 N+ 2}}- 
2 N\, \Lambda r_h^{2 N-1} \Big)\,.
$$
We can eliminate $\Lambda$ from this expression by taking into account that (\ref{sol-eom-AM}) implies, 
at the horizon, 
\beq
P_N(0) = \Lambda r_h^{2 N} + \frac{M}{r_h^{2 d_0 -2 N+ 1}}\,,
\label{implM}
\eeq
(Note incidentally that $M(r_h)$ is trivially obtained from (\ref{implM}). Note also that the value 
$P_N(0)=C[N,d_0]$ is obtained from (\ref{polin}).)

Using (\ref{implM}) we obtain
\beq
T =\frac{1}{4\pi|P_N'(0)|} \Big((2 d_0 + 1) \frac{M}{r_h^{2 d_0 -2 N+ 2}}- \frac{2N}{r_h} P_N(0) \Big)\,,
\label{temper}
\eeq
which in the critical Lovelock dimension, 
$d=2N+2$, gives  
$$
 T_{\!{}_{(d_0=N)}} =\frac{1}{4\pi|P_N'(0)|} \Big((2 N + 1) \frac{M}{r_h^2}- \frac{2N}{r_h} P_N(0) \Big)\,,
$$
which agrees, up to numerical factors, with the results obtained 
in \cite{dpp1} for the spherical pure Lovelock BH.

\vspace{4mm}

\subsection{Stability}

Here we follow the approach of Refs \cite{Maeda:2011ii,Maeda}.
Local thermodynamical stability will correspond to 
positive specific heat, $\displaystyle C_e=\frac{d M}{d T}$. It is easy to compute it as
$$ 
C_e = \frac{M'(r_h)}{T'(r_h)}\,.
$$ 
Using Eq. (\ref{implM}) to express $M$ in terms of $r_h$, we have
\beq
M(r_h)= ( P_N(0) -\Lambda\, r_h^{2N})r_h^{2 d_0-2 N+1}.
\label{MofR}
\eeq
Then temperature in Eq. (\ref{temper}) is given by 
\beq
T =\frac{1}{4\pi|P_N'(0)|} \Big(( 2 d_0 -2 N+ 1) \frac{P_N(0)}{r_h}  - 
( 2 d_0+ 1) \Lambda\,  r_h^{2 N-1}\Big)\,.
\label{tempnoM}\eeq
To keep $M>0$, as required from the analysis in Sec. \ref{partcases},
\beq
P_N(0) > \Lambda\, r_h^{2N}.
\label{LimfromM}
\eeq
Note that $P_N(0)\geq 0$ always. It only vanishes at the critical $d_0=N$ dimension for odd $N$. In all other cases $P_N(0)> 0$. To keep $T>0$, we must have 
\beq
P_N(0) > \frac{( 2 d_0+ 1)}{( 2 d_0 -2 N+ 1)}\Lambda\, r_h^{2N}.
\label{LimfromT}
\eeq
Thus for $\Lambda<0$, the above constraints (\ref{LimfromM}) and (\ref{LimfromT}) impose no 
restrictions whereas 
for $\Lambda>0$ the constraint (\ref{LimfromT}), which anyway implies (\ref{LimfromM}), must be 
satisfied.

From Eq. (\ref{MprimT}) it is clear that $T>0 \Leftrightarrow M'>0$. To have local 
thermodynamical stability we therefore need $T'(r_h)>0$. In fact, since
$$
T'(r_h) = \frac{1}{4\pi|P_N'(0)|} \Big(-( 2 d_0 -2 N+ 1) \frac{P_N(0)}{r_h^2}  - 
( 2 d_0+ 1)(2 N-1) \Lambda\,  r_h^{2 N-2}\Big)\,,
$$
the condition $T'(r_h)>0$ is equivalent to 
\beq
P_N(0) < -\frac{( 2 d_0+ 1)(2 N-1)}{2 d_0 -2 N+ 1}\Lambda\, r_h^{2N}\,.
\label{LimfromTprim}
\eeq
Local thermodynamical stability is thus guaranteed if the bounds (\ref{LimfromT}) and 
(\ref{LimfromTprim}) are satisfied. Since $P_N(0)\geq 0$, Eq.(\ref{LimfromTprim}) 
is inconsistent for positive $\Lambda \geq0$, and hence it is ruled out. 
Thus we need $\Lambda<0$;i.e. $N$ odd. If in addition $d_0=N$, $P_N(0) =0$ and therefore no 
restrictions of any kind. 
If $P_N(0) >0$, there is a lower bound (Eq.(\ref{LimfromTprim})) on BH size, 
\beq
r_h^{2 N} > \frac{2 d_0 -2 N+ 1}{( 2 d_0+ 1)(2 N-1)\vert \Lambda \vert} P_N(0)\,,
\label{LimfromTprim2}
\eeq 
consequently its mass, as given in Eq.(\ref{MofR}), is also bounded from below.  

\vspace{4mm}

Next we consider global thermodynamical stability by requiring free energy to be negative.
Using Eqs (\ref{entr}) and (\ref{tempnoM}), we compute free energy, 
\beq
F = M-T S = \frac{1}{2 d_0 -2 N+ 2}
\Big( P_N(0) r_h^{2 d_0-2 N+1} + (2 N-1) \Lambda \,  r_h^{2 d_0+1}   \Big)\,.
\label{freeE}
\eeq
For $d_0=N$, $P_N(0)=0$ and $\Lambda<0$ ($N$ odd), hence free energy is always 
negative ensuring global stability while for even $N$ and $\Lambda>0$ it is always positive thereby 
implying instability. That is black hole is thermodynamically unstable for $\Lambda>0$ and $N$ even. 
Thus thermodynamical stability requires $\Lambda<0$ and $N$ odd. When $d_0=N$ (pure Lovelock), 
there is no restriction 
of any kind while for $d_0>N$ for which $P_N(0)>0$, there is lower bound on horizon size given by  \\ 
\beq
r_h^{2N} >  \frac{P_N(0)}{(2 N-1)\vert\Lambda\vert}\,,
\label{limit-global}
\eeq
which already implies, as expected, local thermodynamical stability Eq.(\ref{LimfromTprim2}).

Note however that our analysis is formal in the following sense. When $d_0 > N$ the configuration 
with $M=0$ is not allowed in our formalism as it is forbidden from equations (\ref{MofR}) and 
(\ref{limit-global}), since $P_N(0)>0$ . One could think that in the critical, odd $d_0 = N$ case, 
the limit $r_h\to 0$ which sends also $M$ to zero, could be taken without restrictions. And in fact 
the entropy (\ref{entr}) and the temperature (\ref{tempnoM}) (with $P_N(0)=0$) vanish in this limit. 
But then a naked singularity appears at the origin $r=0$ (see eq.(\ref{K}) in section \ref{case3}) 
and thus we must keep $M>0$ to prevent this singularity from forming. This means that the 
global stability analysis must be considered formal. 

\section{Discussion}

We have shown, within the framework of Lovelock theories, that the property that the EOM are polynomial, 
which it is well known to hold in the case of spherically symmetric horizons, can be generalized to 
horizons that are product of two equal spheres (though now the Weyl tensor is not vanishing), 
which are a particular case of Einstein manifolds. 
On the other hand we also know that in the case of Gauss-Bonnet gravity 
(Lovelock N=2) this polynomial property of the EOM holds whenever the horizon is an Einstein 
manifold \cite{Dotti1} (thus including the one-sphere, two-spheres cases). It can be conjectured 
that, in a general Lovelock theory, the polynomial property that we find for two-spheres topology 
could be extended to the more general case of Einstein horizons.
An indication in this direction is the study of $N=3$ Lovelock in \cite{Farhangkhah:2014zka}.

What does two sphere topology entail is that horizon space is non-maximally
symmetric Einstein space which has non-zero Weyl curvature, though
both Weyl and Riemann have vanishing covariant derivative. Product of spheres cause solid angle 
deficits which cancel out each-other's contribution to Ricci tensor \cite{Kol:2002xz} but not to 
Riemann tensor. Thus for $N\geq2$, EOM always has Riemann present and hence contribution to it 
from solid angle deficit has to be anulled out, and that is done with an additional numerical 
parameter $p$, as shown in the $N=2$ case, in which it appears under the radical in the solution. 
It is in fact a measure of non-zero Weyl 
curvature \cite{Dotti1} which is absent for $N=1$. The general feature of vacuum equation, 
ultimately reducing to an algebraic polynomial is carried through even for two sphere topology 
for $d_0\geq N$. That is why 
we have been able to solve pure Lovelock vacuum equation for arbitrary order $N$ in this enlarged 
general setting. Though our general analysis refers to the critical dimension $d_0=N$, 
however the equation reducing to an algebraic polynomial is a general feature true for all $d_0\geq N$.  

It turns out that in the critical $d_0=N$ case, constant $p$ and $\Lambda$ always bear opposite sign, 
the former is positive or negative according as $N$ is odd or even and opposite is the case for $\Lambda$. 
For even $N$, 
$p<0$ and $\Lambda>0$, spacetime has two horizons, black  hole and cosmological, and black hole mass 
is constrained in a range given in terms of $\Lambda$. While for odd $N$, $\Lambda<0$, there is only 
one black hole horizon without any restriction. Remarkably thermodynamical stability also has the 
similar discerning feature. That is spacetime is thermodynamically stable for odd $N$ and $\Lambda<0$ 
while it is unstable for even $N$ and $\Lambda>0$. 

We thus conclude that two sphere topology for pure Lovelock gravity clearly discerns between odd and 
even Lovelock order $N$, consequently negative and positive $\Lambda$ and it is the former which is 
favored. 
This is a remarkable and interesting property of this setting which has been revealed only when we 
probed higher orders, $N>2$. Let's reiterate that universality of kinematic property of gravity 
in all odd $d=2N+1$ dimensions demands the pure Lovelock equation which is valid only for the two 
$d=2N+1, 2N+2$ dimensions. This is the main and critical motivation for pure Lovelock gravity. 
In the context of the present study we therefore always take $d=2N+2$ which implies $d_0=N$.

\section*{Acknowledgement} This work has been supported through grants FPA2013-46570, 2014-SGR-104
and Consolider CPAN. ND wishes to acknowledge Departament ECM Universitat de Barcelona for a kind 
invitation for a visit that facilitated this work. We also thank the Referee for constructive 
criticism and useful suggestions.

\section*{Appendix A: Proof that $[{\cal L}_N]_0$ is a derivative}
\label{proof}
Consider equation (\ref{lovelockEOM}) for $\rho=\mu=0$. The factor $\sqrt{-g}$ is, up to irrelevant 
dependences on the angles, $r^{2 d_0}$. Obviously the term with the cosmological constant is an 
$r$-derivative, so we will only consider in this appendix the expression $[{\cal L}_N]_0$ with 
vanishing $\Lambda$, 
\beq
[{\cal L}_N]_0=r^{2 d_0}\, \delta^{0\,\mu_1\dots \mu_{N}\nu_1\dots \nu_N}_{0\,\rho_1\dots \rho_{N}
\sigma_1\dots \sigma_N} \, 
 R_{\mu_1\nu_1} \!{}^{\!\rho_1\sigma_1}\dots  R_{\mu_N\nu_N}\! {}^{\!\rho_N \sigma_N}\,.
 \label{lovelockEOM1}
\eeq
This expresion is a sum of products of $N$ components of the Riemann tensor. 
We must fill in it the components (\ref{theDterms}). Since the index $0$ will not be present in 
these components, we can only introduce the factors  $D(1,i), D(1,i'), D(i,j), D(i',j'), D(i,i')$. 
Note that the first two objects in this list contain $A'(r)$, whereas the rest only contain $A(r)$. 
So (\ref{lovelockEOM1}) will become the sum of two objects, one of them containing $A'(r)$ 
as a factor and the 
other without this derivative. To construct each of these objects one must consider all possible 
distributions of the factors $D$ already mentioned, which is a matter of combinatorics. 
In the procedure one must take into account the location of each $D$ within any of the $N$ 
slots available in each additive term and, additionally, the combinatorics associated with the 
choice of angles, $i,\, i'$, for each $D$.

All in all, using the quantities (\ref{theC}) and (\ref{polin}), and defining the new ones,
$$
\bar C[m,s] = \frac{1}{2^m} \sum_{k=0}^m {m\choose k} \frac{s !}{(s-2 m + 2 k) !} 
\frac{(s-1) !}{(s- 2 k -1) !} \,,
$$
$$
 F_N(A,A')=N \sum _{l=0}^{N-1} \Big(\frac{d_0 !}{(d_0-l)!}\Big)^2 (d_0-l)\, \bar C[N-l-1,d_0-l]\,
 \,(-A')(-A)^l (1-A)^{N-l-1}\,,
$$
\Big( An alternative form for $F_N(A,A')$ is:
\bea
F_N(A,A')&=&\sum _{l=0}^{N-1} \sum _{k=0}^{N-l-1}\frac{1}{2^{N-l-1}}\,\frac{N!}{k! \,l!\,(N-k-l-1)!}\,
\frac{d_0!}{(d_0-2 k-l-1)!}\nonumber\\
&& \frac{d_0!}{(d_0+2 k-2
   (-l+N-1)-l)!}
(-A'(r))(-A(r))^l  (1-A(r))^{N-l-1}\nonumber \  \Big)
\eea
we end up with the result
$$
\frac{1}{4^N}[{\cal L}_N]_0=r^{2 d_0-2N+1}F_N(A,A')+r^{2 d_0-2N}P_N(A)\,,
$$
but it turns out that
$$
\frac{d}{d r}P_N(A) = (2 d_0-2N+1)F_N(A,A')\,,
$$
and therefore
$$
\frac{1}{4^N}[{\cal L}_N]_0=
\frac{1}{2 d_0-2N+1}\,\frac{d}{d r}\Big(r^{2 d_0-2N+1}P_N(A)\Big)\,,
$$
which proves our claim.

\section*{Appendix B: Proof of (\ref{app})}
Here we prove (\ref{app}). To this effect we will consider the Noether identities for gauge theories 
applied to the Lovelock Lagrangians. These Lagrangians are generally covariant -and hence gauge theories- 
under the diffeomorphism group (coordinate reparametrizations). 

\vspace{4mm}

Consider a field theory with dynamics derived from a variational principle with Lagrangian ${\cal L}$ 
and fields $\phi^A$ (here $\phi^A$ means a generic field or field component). The Noether 
symmetries (satisfying that $\delta{\cal L}$ is a divergence) 
are implemented by infinitesimal transformations
\beq
\delta \phi^A = R^A_a \epsilon^a + R^{A\mu}_a \partial_\mu\epsilon^a\,,
\label{gaugetr}
\eeq
with $\epsilon^a$ being the infinitesimal parameters of the symmetries, and with the index $a$ running 
over the number of independent symmetries. The coefficients $R^A_a,\ R^{A\mu}_a$ are functions of the 
fields and their derivatives. The {\sl gauge} character of the symmetries  relies in the fact that the 
functions $\epsilon^a$ are {\sl arbitrary} functions of the coordinates -and even of the fields. 
If $[{\cal L}]_A$ are the functional (Euler-Lagrange) derivatives with respect to $\phi^A$, 
the Noether identities are written as
\beq
[{\cal L}]_A R^A_a-\partial_\mu([{\cal L}]_A R^{A\mu}_a)=0\,.
\label{noethid}
\eeq
Applying these identities to the Einstein-Hilbert 
Lagrangian one obtains the well known doubly contracted Bianchi identities for the Riemann tensor.
\vspace{4mm}

For convenience, we will move momentarily to the formalism that uses the vielbein and the Lorentz 
connection as the independent fields. The vielbein $e^{\,I}_\mu$ extracted from the metric 
(\ref{configA}) is (we use the notation $I=\u{\nu}$ for the flat indices)
\beq
e^{\,\u{0}}_0=\sqrt{A(r)},\qquad e^{\,\u{1}}_1=\frac{1}{\sqrt{A(r)}},\qquad 
e^{\,\u{i}}_i=r f_i,\qquad e^{\,\u{i'}}_{i'}=r f_{i'},
\label{vielb}
\eeq
where $f_i$ and $f_{i'}$ are trigonometric functions of the angles of the spheres whose specific 
form is irrelevant to us. notice that the vielbein is ``diagonal'' in its indices.

The Lovelock Lagrangian (with the cosmological constant term) in this formalism will be denoted 
$\tilde{\cal L}_N$ and its EOM are the 
functional derivatives with respect to the vielbein and the Lorentz connection. 
The functional derivatives with respect to the connection are 
proportional to the torsion tensor and we require this tensor to vanish in order to recover our original setting where 
the connection was just the Levi Civita connection. Thus we need not worry about the EOM coming from 
variations of the connection.

There are two types of gauge symmetries in the present formalism, namely diffeomorphism invariance and
local Lorentz invariance. The Noether identities for the local Lorentz invariance are empty in our case. 
So we will focus on diffeomorphism invariance. The vielbein vectors  $e^{\,I}_\mu$ transform, under
a difeomorphism $\delta x^\mu =- \epsilon^\mu$, with the Lie derivative
$$
\delta e^{\,I}_\mu = \epsilon^\rho \partial_\rho e^{\,I}_\mu + e^{\,I}_\rho\partial_\mu\epsilon^\rho\,,
$$
and thus we infer from (\ref{gaugetr}) the identifications
$$
R^{(e^{\,I}_\mu)}_\rho = \partial_\rho e^{\,I}_\mu,\qquad 
R^{(e^{\,I}_\mu)\,\sigma}_\rho= \delta^{\,\sigma}_\mu e^{\,I}_\rho\,.
$$
For the N-Lovelock Lagrangian the Noether identities (\ref{noethid}) become
\beq
[\tilde{\cal L}_N]_{(e^{\,I}_\mu)}\,R^{(e^{\,I}_\mu)}_\rho -
\partial_\sigma([\tilde{\cal L}_N]_{(e^{\,I}_\mu)}\, R^{(e^{\,I}_\mu)\,\sigma}_\rho)=0\,.
\label{noethidLov}
\eeq

As an identity, (\ref{noethidLov}) holds for any field configuration, and in particular for 
(\ref{vielb}) (which implies (\ref{configA})). When we plug (\ref{vielb}) into (\ref{noethidLov}) 
we can ease the notation because, the vielbein being ``diagonal'', each contributing field is 
identified by a $\mu$ index: $e^{\,\u{\mu}}_\mu\rightarrow \mu$, we get
\beq
[\tilde{\cal L}_N]_\mu\,R^{(\mu)}_\rho -
\partial_\sigma([\tilde{\cal L}_N]_\mu\, R^{(\mu)\,\sigma}_\rho)=0\,.
\label{noethidLov2}
\eeq
It is easy to see that the functional derivatives $[\tilde{\cal L}_N]_\mu$ are proportional to those in 
(\ref{lovelockEOM}). To be specific, the result is 
\beq
[\tilde{\cal L}_N]_\mu = \frac{1}{e^{\,\u \mu}_\mu}[{\cal L}_N]_\mu\, 
\label{tild}
\eeq
(clearly no sum for $\mu$).

The only nontrivial case for (\ref{noethidLov2}) is when $\rho=1$ (the radial variable). We obtain,
using (\ref{eq0}), (\ref{eqi}) and (\ref{tild}),
\bea
0 &=&[\tilde{\cal L}_N]_\mu\,\partial_1 e^{\,\u \mu}_\mu-
\partial_\mu([\tilde{\cal L}_N]_\mu\,e^{\,\u \mu} _1)=[\tilde{\cal L}_N]_\mu\,\partial_1 e^{\,\u \mu}_\mu-
\partial_1([\tilde{\cal L}_N]_1\,e^{\,\u 1} _1)\\ &=&
[\tilde{\cal L}_N]_i\,\partial_1 e^{\,\u i}_i+[\tilde{\cal L}_N]_{i'}\,\partial_1 e^{\,\u i'}_{i'} 
+[\tilde{\cal L}_N]_0\,\partial_1 e^{\,\u 0}_0+[\tilde{\cal L}_N]_{1}\,\partial_1 e^{\,\u 1}_{1} -
\partial_1(\tilde[{\cal L}_N]_1\,e^{\,\u 1} _1)\\ &=&
[\tilde{\cal L}_N]_i\,f_i+[\tilde{\cal L}_N]_{i'}\,f_{i'} +[\tilde{\cal L}_N]_0\,\partial_1 e^{\,\u 0}_0
 -\partial_1(\tilde[{\cal L}_N]_1)\,e^{\,\u 1} _1\\
&=& \frac{2 d_0}{r} [{\cal L}_N]_{i} +
\frac{1}{\sqrt{A(r)}}[{\cal L}_N]_0\, \partial_1\sqrt{A(r)}- 
\partial_1 ([{\cal L}_N]_1 \sqrt{A(r)})\frac{1}{\sqrt{A(r)}}\\
&=& \frac{2 d_0}{r} [{\cal L}_N]_{i} - \partial_1 ([{\cal L}_N]_0)\,.
\label{noethidLov3}
\eea
which is the result (\ref{app}).


\begin{thebibliography}{90}

\bibitem{Dadhich:2012cv}
  N.~Dadhich, S.~G.~Ghosh and S.~Jhingan,
  ``The Lovelock gravity in the critical spacetime dimension,''
  Phys.\ Lett.\ B {\bf 711} (2012) 196
  [arXiv:1202.4575 [gr-qc]].
\bibitem{Dadhich:2008df}
  N.~Dadhich,
  ``Characterization of the Lovelock gravity by Bianchi derivative,''
  Pramana {\bf 74} (2010) 875
  [arXiv:0802.3034 [gr-qc]].


  
\bibitem{BD}
  D.~G.~Boulware and S.~Deser,
 ``String Generated Gravity Models,''
  Phys.\ Rev.\ Lett.\  {\bf 55} (1985) 2656.
  
\bibitem{Wheeler:1985qd}
  J.~T.~Wheeler,
  ``Symmetric Solutions to the Maximally {Gauss-Bonnet} Extended Einstein Equations,''
  Nucl.\ Phys.\ B {\bf 273} (1986) 732.
  
  

\bibitem{Wheeler:1985nh}
  J.~T.~Wheeler,
  ``Symmetric Solutions to the Gauss-Bonnet Extended Einstein Equations,''
  Nucl.\ Phys.\ B {\bf 268} (1986) 737.

\bibitem{Whitt:1988ax}
  B.~Whitt,
  ``Spherically Symmetric Solutions of General Second Order Gravity,''
  Phys.\ Rev.\ D {\bf 38} (1988) 3000.
  
\bibitem{Banados:1993ur}
  M.~Banados, C.~Teitelboim and J.~Zanelli,
  ``Dimensionally continued black holes,''
  Phys.\ Rev.\ D {\bf 49} (1994) 975
  [gr-qc/9307033].
  
  
  
\bibitem{Crisostomo:2000bb}
  J.~Crisostomo, R.~Troncoso and J.~Zanelli,
  ``Black hole scan,''
  Phys.\ Rev.\ D {\bf 62} (2000) 084013
  [hep-th/0003271].
  
  

\bibitem{dpp1}
N.~Dadhich, J.~M.~Pons and K.~Prabhu,
  ``Thermodynamical universality of the Lovelock black holes,''
  Gen.\ Rel.\ Grav.\  {\bf 44} (2012) 2595
  [arXiv:1110.0673 [gr-qc]].
  
  
\bibitem{Camanho:2011rj} 
  X.~O.~Camanho and J.~D.~Edelstein,
  ``A Lovelock black hole bestiary,''
  Class.\ Quant.\ Grav.\  {\bf 30}, 035009 (2013)
  [arXiv:1103.3669 [hep-th]].
  
  
\bibitem{Maeda:2011ii}
  H.~Maeda, S.~Willison and S.~Ray,
  ``Lovelock black holes with maximally symmetric horizons,''
  Class.\ Quant.\ Grav.\  {\bf 28} (2011) 165005
  [arXiv:1103.4184 [gr-qc]].
  
  \bibitem{nariai}
H.~Nariai, Sci. Rep. Tohoku Univ. {\bf 34} (1950) 160; {\bf 35} (1951) 62.

  
\bibitem{Dotti1}
  G.~Dotti and R.~J.~Gleiser,
  ``Obstructions on the horizon geometry from string theory corrections to Einstein gravity,''
  Phys.\ Lett.\ B {\bf 627} (2005) 174
  [hep-th/0508118].

\bibitem{Zegers}
  C.~Bogdanos, C.~Charmousis, B.~Gouteraux and R.~Zegers,
  ``Einstein-Gauss-Bonnet metrics: Black holes, black strings and a staticity theorem,''
  JHEP {\bf 0910} (2009) 037
  [arXiv:0906.4953 [hep-th]].
  
\bibitem{Dotti:2008pp}
  G.~Dotti, J.~Oliva and R.~Troncoso,
  ``Vacuum solutions with nontrivial boundaries for the Einstein-Gauss-Bonnet theory,''
  Int.\ J.\ Mod.\ Phys.\ A {\bf 24} (2009) 1690
  [arXiv:0809.4378 [hep-th]].
\bibitem{Dotti2}
G.~Dotti, J.~Oliva and R.~Troncoso,
 ``Static solutions with nontrivial boundaries for the Einstein-Gauss-Bonnet theory in vacuum,''
  Phys.\ Rev.\ D {\bf 82} (2010) 024002
  [arXiv:1004.5287 [hep-th]].
  
  \bibitem{Maeda}
  H.~Maeda,
  ``Gauss-Bonnet black holes with non-constant curvature horizons,''
  Phys.\ Rev.\ D {\bf 81} (2010) 124007
  [arXiv:1004.0917 [gr-qc]].
  
\bibitem{Pons:2014oya}
  J.~M.~Pons and N.~Dadhich,
  ``On static black holes solutions in Einstein and Einstein-Gauss-Bonnet gravity with topology ${\bf SO(n) \times SO(n)}$,''
  arXiv:1408.6754 [gr-qc].
\bibitem{dpp2}
N.~Dadhich, J.~M.~Pons and K.~Prabhu, 
  "On the static Lovelock black holes," 
  Gen. Relativ. Grav. {\bf 45} (2013) 1131  
  [arXiv:1201.4994 [gr-qc]].
 
\bibitem{Kol:2002xz} 
  B.~Kol,
  ``Topology change in general relativity, and the black hole black string transition,''
  JHEP {\bf 0510}, 049 (2005)
  [hep-th/0206220].
  
\bibitem{Farhangkhah:2014zka} 
  N.~Farhangkhah and M.~H.~Dehghani,
 ``Lovelock black holes with nonmaximally symmetric horizons,''
  Phys.\ Rev.\ D {\bf 90}, no. 4, 044014 (2014)
  [arXiv:1409.1410 [gr-qc]].

\end{thebibliography}
\end{document}